\documentclass[preprint, review, 12pt, 3p]{elsarticle}

\usepackage{amsmath}
\usepackage{amssymb}
\usepackage[hidelinks]{hyperref}
\usepackage{graphicx}
\usepackage{subfig}
\usepackage{multirow}
\usepackage{lipsum}
\usepackage{makecell}
\usepackage{dirtytalk}
\usepackage{natbib}
\usepackage{caption}
\usepackage{adjustbox}
\usepackage{tabularx}
\usepackage{floatrow} 
\usepackage[vlines]{tabularht}
\usepackage{pbox}
\usepackage{rotating}
\usepackage{longtable,lscape}
\usepackage{xltabular}
\usepackage{multirow}
\usepackage{colortbl}
\usepackage{hhline}
\usepackage{hyperref}
\usepackage{float}
\floatstyle{plaintop}
\restylefloat{table}

\usepackage[linesnumbered,ruled,vlined]{algorithm2e}
\SetKwInput{KwInput}{Input}                
\SetKwInput{KwOutput}{Output}              

\usepackage[table,xcdraw]{xcolor}

\newsavebox\MBox

\biboptions{sort&compress, square}

\journal{Informatics in Medicine Unlocked}

\begin{document}

\begin{frontmatter}

\title{Acute Lymphoblastic Leukemia Detection from Microscopic Images Using Weighted Ensemble of Convolutional Neural Networks}

\author[label1,label4]{Chayan Mondal}
\ead{chayan.eee@bsmrstu.edu.bd}

\author[label1,label5]{Md. Kamrul Hasan\corref{cor1}}
\ead{m.k.hasan@eee.kuet.ac.bd}

\author[label1]{Md. Tasnim Jawad}
\ead{jawad1703006@stud.kuet.ac.bd}

\author[label2]{Aishwariya Dutta}
\ead{aishwariyadutta16@gmail.com}

\author[label1]{Md. Rabiul Islam}
\ead{rabiulnewemail@gmail.com}

\author[label3]{Md. Abdul Awal}
\ead{m.awal@ece.ku.ac.bd}

\author[label1]{Mohiuddin Ahmad}
\ead{ahmad@eee.kuet.ac.bd}

\address[label1]{Department of Electrical and Electronic Engineering (EEE), Khulna University of Engineering \& Technology, Khulna-9203, Bangladesh}

\address[label2]{Department of Biomedical  Engineering (BME), Khulna University of Engineering \& Technology, Khulna-9203, Bangladesh}

\address[label3]{Electronics and Communication Engineering (ECE) Discipline, Khulna University,
Khulna-9208, Bangladesh}

\address[label4]{Department of Electrical and Electronic Engineering (EEE), Bangabandhu Sheikh Mujibur Rahman Science \& Technology University, Gopalganj-8100, Bangladesh}

\cortext[cor1]{I am corresponding author}
\fntext[label5]{Department of EEE, KUET, Khulna-9203, Bangladesh.}



\begin{abstract}
Acute Lymphoblastic Leukemia (ALL) is a blood cell cancer characterized by numerous immature lymphocytes. Even though automation in ALL prognosis is an essential aspect of cancer diagnosis, it is challenging due to the morphological correlation between malignant and normal cells. The traditional ALL classification strategy demands experienced pathologists to carefully read the cell images, which is arduous, time-consuming, and often suffers inter-observer variations. This article has automated the ALL detection task from microscopic cell images, employing deep Convolutional Neural Networks (CNNs). We explore the weighted ensemble of different deep CNNs to recommend a better ALL cell classifier. The weights for the ensemble candidate models are estimated from their corresponding metrics, such as accuracy, F1-score, AUC, and kappa values. Various data augmentations and pre-processing are incorporated for achieving a better generalization of the network. We utilize the publicly available C-NMC-2019 ALL dataset to conduct all the comprehensive experiments. Our proposed weighted ensemble model, using the kappa values of the ensemble candidates as their weights, has outputted a weighted F1-score of $88.6\,\%$, a balanced accuracy of $86.2\,\%$, and an AUC of $0.941$ in the preliminary test set. The qualitative results displaying the gradient class activation maps confirm that the introduced model has a concentrated learned region. 
In contrast, the ensemble candidate models, such as Xception, VGG-16, DenseNet-121, MobileNet, and InceptionResNet-V2, separately produce coarse and scatter learned areas for most example cases. Since the proposed kappa value-based weighted ensemble yields a better result for the aimed task in this article, it can experiment in other domains of medical diagnostic applications.
\end{abstract}


\end{frontmatter}


\section{Introduction}
\label{introduction}

\subsection{Problem presentation}
\label{Problempresentation}
Cancer, a group of uncommon and distinctive diseases, is one of the deadliest diseases \citep{gehlot2020sdct}, which is abnormal and uncontrolled cell growth. 
In 2020, World Health Organization (WHO) claimed that approximately 19.3 million people were diagnosed with cancer, caused a death of 10 million people, which is almost 1.6 times greater than in 2000 \citep{WHO}.
The affected number is expected to be around 50 percent higher in 2040 than now \citep{WHO}. 
Among various types of cancer, one of the most common types of childhood cancer is Acute Lymphoblastic Leukemia (ALL), which affects the White Blood Cells (WBCs) \citep{CancerLeukemia}. 
ALL patients have an excessive amount of premature WBCs in their bone marrow and can spread to other organs, like the spleen, liver, lymph nodes, central nervous system, and testicles \citep{ACSIntro}. 
Although the leading causes of ALL are unknown yet, several representatives, like contact with severe radiation and chemicals, such as benzene and infection with T-cell lymphoma, can boost the possibility of generating ALL  \citep{cancerorg}. 
Almost $55.0\,\%$ of total worldwide ALL cases are caused in the Asia Pacific region \citep{solomon2017global}. 
According to WHO, ALL's total cases are 57377, which is $21.9\,\%$ of the worldwide total childhood cancer cases in $2020$ \citep{WHO2020}.

Generally, the doctors suspect ALL patients through specific symptoms and signs, where different clinical inspections authenticate the ALL diagnosis \citep{ACS}.  
The blood examinations are frequently performed on the suspected ALL patients in the preliminary stage. The complete blood count and peripheral blood smear inspections are accomplished to monitor the changes in the numbers and appearances of WBC in blood cells, respectively \citep{ACS}. 
The diagnosis of ALL with higher accuracy is achieved by utilizing chromosome-based tests, such as cytogenetics, fluorescent in situ hybridization, and polymerase chain reaction, where chromosomes are observed to recognize unusual blood cells \citep{ACS}.
An image-based automated Computer-aided Prognosis (CAP) tool with negligible false-negative rates is a crying requirement to accelerate ALL patients' diagnosis in early stages, as the survival rate is as high as $90.0\,\%$ with early detection \citep{STJUDE}.
Different image-based prognoses are being applied to diagnose ALL patients \citep{ACS}, utilizing Computed Tomography (CT) \& Magnetic Resonance Imaging (MRI) scans, X-rays, and Ultrasound (US). 
However, the collections of those imaging modalities are costly and time-consuming, requiring an expert pathologist, or hematologist, or oncologist \citep{gehlot2020sdct}. 
Moreover, the scanners of those images are still unavailable in under-developed and developing countries, especially in rural regions, according to a report published by WHO in 2020 \citep{WHO2020Global}.
Currently, a microscopic image-based CAP system for ALL analysis can overcome these limitations because these can be fully automated and do not require highly trained medical professionals to run the tests \citep{gehlot2020sdct}. 
In the last ten years, the efficiency of Deep Learning (DL)-based methods for automating CAP systems increased dramatically, and their performances seem to outperform conventional image processing methods in image classification tasks \citep{asiri2019deep}.
However, the DL-based strategies have superior reproducibility than the Machine Learning (ML)-based approaches; the latter methods require handcrafted feature engineering \citep{mishra2019texture}. 
Different DL-based methods have already proven their tremendous success in varying fields of automatic classification, detection, or segmentation, such as skin lesion \citep{hasan2021dermo, hasan2021dermoexpert, hasan2020dsnet, dutta2020skin}, breast cancer \citep{9230708, steiner2018impact}, brain tumor \citep{8858515, icsin2016review}, diabetic retinopathy \citep{hasan2021drnet}, COVID-19 pandemic \citep{hasan2021covid, hasan2020challenges, oh2020deep, lalmuanawma2020applications}, minimally invasive surgery \citep{hasan2021detection}, and others \citep{7873105}. 
This article will explore and perform an in-depth study of the value of DL methods for the image-based ALL prognoses. Different ways of ALL predictions are briefly reviewed in the following section.

\subsection{Literature Review}
\label{Literature}
This section presents the review of current CAP methods for the analysis of ALL, where we first discuss the ML-based systems, then subsequently DL-based approaches.  

\paragraph{\textbf{ML-based methods}}
\citet{Mohapatra2011} proposed a fuzzy-based color segmentation method to segregate leukocytes from other blood components, followed by the nucleus shape and texture extraction as discriminative features. Finally, the authors applied a Support Vector Machine (SVM) \citep{furey2000support} to detect leukemia in the blood cells.
The k-means Clustering (KMC)-based segmentation \citep{macqueen1967some} was employed by \citet{madhukar2012new} to extract the leukocytes' nuclei using color-based clustering. Different types of features, such as shape (area, perimeter, compactness, solidity, eccentricity, elongation, form-factor), GLCM \citep{ondimu2008effect} (energy, contrast, entropy, correlation), and fractal dimension were extracted from the segmented images. Finally, they applied the SVM classifier, utilizing K-fold, Hold-out, and Leave-one-out cross-validation techniques. 
\citet{joshi2013white} developed a blood slide-image segmentation method followed by a feature extraction (area, perimeter, circularity, etc.) policy for detecting leukemia. The authors utilized the k-Nearest Neighbor (KNN) \citep{hasan2017prediction} classifier to classify lymphocyte cells as blast cells from normal white blood cells.
\citet{mishra2019texture} proposed a discrete orthonormal S-transform \citep{stockwell2007basis}-based feature extraction followed by a hybrid Principal Component Analysis (PCA) and linear discriminant analysis-based feature reduction approach for a lymphoblastic classification scheme. Finally, the author classified those reduced features using an AdaBoost-based Random Forest (ADBRF) \citep{hasan2020diabetes} classifier.
The authors in \citep{mahmood2020identification} aimed at four machine learning-based algorithms, such as classification and regression trees (CART), RF, Gradient Boosted (GM) engine \citep{greenwell2019gbm}, and C$5.0$ decision tree algorithm \citep{kuhn2014c50}. Their experiment demonstrated the superior performance of the CART method. 
\citet{fathi2020design} produced an integrated approach combining PCA, neuro-fuzzy, and GMDH (group method of data handling) to diagnose ALL, which helps to detect two types of leukemia, such as ALL and acute myeloid leukemia.
\citet{kashef2020treatment} recommended different ML algorithms, such as decision tree \citep{hasan2020diabetes}, SVM, linear discriminant analysis, multinomial linear regression, gradient boosting machine, RF, and XGBoost \citep{hasan2020diabetes}, where the XGBoost algorithm exhibited the best results. 
Authors in \citep{gebremeskel2021automatic} developed a K-means image segmentation and marker controlled segmentation-based classification and detection algorithms, where multi-class SVM was used as a classifier. 
Table~\ref{tab:summary_ML_methods} shows a summary of several ML-based models for ALL classification with their respective pre-processing, utilized datasets, and classification results in terms of accuracy.
\begin{table*}[!ht]
\caption{Summary ML-based methods for ALL classification, including publication year, pre-processing \& classification techniques, used datasets, and corresponding results in accuracy (Acc).}
\centering
\scriptsize
\begin{tabular}{lp{4cm}p{4.4cm}p{1.3cm}p{2.2cm}l}
\hline
  {\cellcolor[HTML]{C0C0C0}Years} &
  {\cellcolor[HTML]{C0C0C0}Pre-processing} & {\cellcolor[HTML]{C0C0C0} Features} &
  {\cellcolor[HTML]{C0C0C0}Classifier} &
  {\cellcolor[HTML]{C0C0C0}Datasets} &
  {\cellcolor[HTML]{C0C0C0}Acc}
 \\ \hline
$2011$ \citep{Mohapatra2011} & Median filtering \& unsharp masking & Hausdorff dimension, contour signature, fractal dimension, shape, color, and texture features & SVM & ALL-IGH \citep{Mohapatra2011} & $93.0\,\%$ \\  \hline
 
$2012$ \citep{madhukar2012new} & KMC, color correlation, and contrast
enhancement & Shape, GLCM, and fractal dimension features & SVM & ALL-FS \citep{scotti2005automatic}  & $93.5\,\%$ \\  \hline
 
$2012$ \citep{supardi2012classification} & No & Twelve size-, color-, and shape-based features & KNN & ALL-HUSM  \citep{supardi2012classification} & $80.0\,\%$ \\  \hline

$2013$ \citep{joshi2013white} & Threshold-based segmentation & Shape-based (area, perimeter, and circularity) features & KNN & ALL-IDB \citep{labati2011all} & $93.0\,\%$ \\  \hline

$2014$ \citep{laosai2014acute} & Color correction, conversion, and KMC & Texture, geometry, and statistical features & SVM and KMC& ALL-UCH \citep{laosai2014acute} & $92.0\,\%$ \\  \hline
 
$2015$ \citep{viswanathan2015fuzzy} & Contrast enhancement and morphological segmentation  & Texture, geometry, color and statistical features & Fuzzy cluster  & ALL-IDB \citep{labati2011all} & $98.0\,\%$ \\  \hline

$2019$ \citep{mishra2019texture} & Color conversion and thresholding for segmentation & Morphological, textural, and colour-based features  & ADBRF & ALL-IDB1\citep{labati2011all} & $99.7\,\%$ \\  \hline

$2021$ \citep{gebremeskel2021automatic} & Resizing, contrast enhancement, and KMC & Texture, geometry, and color features & SVM & ALL-JUSH \citep{gebremeskel2021automatic} & $94.6\,\%$ \\  \hline
\end{tabular}
\label{tab:summary_ML_methods}
\end{table*}

\paragraph{\textbf{DL-based methods}} 
\citet{honnalgere2019classification} proposed a VGG-16 \citep{simonyan2014very} network, which was fine-tuned with batch normalization and pre-trained on the ImageNet dataset \citep{deng2009imagenet}. A DL-based framework was developed by \citet{marzahl2019classification}, using the normalization-based pre-processing step and two augmentation methods. They used ResNet-18 \citep{he2016deep} network and an additional regression head to predict the bounding box for classification. 
In \citep{shah2019classification}, the authors introduced a DCT-based ensemble model, a combination of Convolutional and Recurrent Neural Networks (CNN-RNN) for the classification of normal versus cancerous cells. In their hybrid model, pre-trained CNN was employed to extract features, whereas the RNN was utilized to extract the dynamic spectral domain features. This ensemble-based model, the combination of DCT-LSTM and its fusion with a pre-trained CNN architecture (AlexNet \citep{krizhevsky2012imagenet}), made this classifier robust and efficient. The pre-processing scheme with crop contour and data augmentation technique increased the aforementioned proposed architecture's accuracy. 
\citet{ding2019deep} presented three different deep learning-based architectures, such as Inception-V3 \citep{szegedy2016rethinking}, DenseNet-121 \citep{huang2017densely}, and InceptionResNet-V2 \citep{szegedy2017inception} for white blood cancer microscopic images classification model. Also, they proposed an ensemble neural network model and demonstrated that their developed stacking model performed better than individually any other single classification model employed in their experiment.
In \citep{shi2019ensemble}, the authors recommended an ensemble of state-of-the-art CNNs (SENet and PNASNet) classification models. They adopted the Grad-CAM technique to scrutinize the CNN model's stability and visualize each cell's most prominent part.
\citet{prellberg2019acute} conferred a leukemia cell classification model using ResNeXt \citep{xie2017aggregated} model with Squeeze-and-Excitation modules \citep{hu2018squeeze}. 
The authors in \citep{kulhalli2019toward} produced an automated stain-normalized white blood cell classifier that can classify a malignant (B-ALL) cell or a healthy (HEM) cell. They used the same ResNeXt ($50$ and $101$) model's ensemble technique and showed that the ResNeXt variants are performed best accordingly. 
\citet{pan2019neighborhood} introduced the Neighborhood-correction Algorithm (NCA) for normal versus malignant cell classification for microscopic images. 
The authors combined ResNet \citep{he2016deep} architecture's advantages (ResNet-50, ResNet-101, ResNet-152) and constructed a fisher vector \citep{sanchez2013image}. According to weighted majority voting, they corrected the initial label of the test cell images.
Authors in \citep{anwar2020convolutional} proposed a ten-layer CNN architecture to detect ALL automatically. 
In \citep{safuan2020investigation}, the authors compared three different deep learning-based algorithms, such as AlexNet, GoogleNet \citep{szegedy2015going}, and VGG classifier model, to detect lymphoblast cells. 
Recently, \citet{gehlot2020sdct} developed the SDCT-AuxNet$\theta$ classifier that uses features from CNN and other auxiliary classifiers. Rather than traditional RGB images, stain deconvolved quantity images were utilized in their work.
Table~\ref{tab:summary_DL_methods} summarizes several DL-based models for ALL classification with their respective pre-processing, utilized datasets, and classification results in terms of F1-score.

\begin{table*}[!ht]
\caption{Summary DL-based methods for ALL classification, including publication year, pre-processing \& classification techniques, used datasets, and corresponding results in F1-score (FS).}
\centering
\scriptsize
\begin{tabular}{lp{5.5cm}p{1cm}p{4cm}p{1.8cm}l}
\hline
  {\cellcolor[HTML]{C0C0C0}Years} &
  {\cellcolor[HTML]{C0C0C0}Pre-processing and augmentations} & {\cellcolor[HTML]{C0C0C0}Features} &
  {\cellcolor[HTML]{C0C0C0}Classifier} &
  {\cellcolor[HTML]{C0C0C0}Datasets} &
  {\cellcolor[HTML]{C0C0C0}FS}
 \\ \hline
  
$2017$ \citep{duggal2017sd} & Normalization, segmentation, random rotations, and vertical flipping
  &  No
  & AlexNet and Texture-based CNN & BRAIRCH \citep{duggal2017sd} & $95.4\,\%$ \\  \hline
  
$2019$ \citep{marzahl2019classification} & Normalization, flipping, rotation, scaling, contrast enhancement, and tilting & No
 & ResNet-18 for classification and detection  & C-NMC \citep{Gupta2019}
 & $87.5\,\%$ \\  \hline
 
$2019$ \citep{ding2019deep} & Center crop, random affine transformation, normalization, rotation, scaling, horizontal, and vertical flipping & No & Ensemble of Inception-V3, Densenet-121, and InceptionResNet-V2 & C-NMC \citep{Gupta2019} & $86.7\,\%$ \\  \hline

$2019$ \citep{shi2019ensemble} & Pixel-wise normalization, randomly resized \& rotated, and center cropping & No & Ensemble of SENet-154 and PNASNet & C-NMC \citep{Gupta2019} & $86.6\,\%$ \\  \hline
 
$2019$ \citep{kulhalli2019toward} & Vertical and horizontal flipping, shearing, distortion, zooming, cropping, and skewing  & No & Different variants of ResNeXt & C-NMC \citep{Gupta2019} & $85.7\,\%$ \\  \hline

$2019$ \citep{Liu2019} & Region segmentation, stain normalization, random flipping, rotation, Gaussian noise addition, contrast, and color adjustment & No & Deep bagging ensemble of Inception and ResNet & C-NMC \citep{Gupta2019} & $84.0\,\%$ \\  \hline

$2019$  \citep{Verma2019} & Center cropping, normalization, and  resizing  & No & MobileNet-V2  & C-NMC \citep{Gupta2019} & $87.0\,\%$ \\  \hline

$2019$ \citep{Khan2019} & Center cropping, random flipping, and rotation & No & Ensemble of ResNet-34, ResNet-50, and ResNet-101 & C-NMC \citep{Gupta2019} & $81.7\,\%$ \\  \hline

$2019$  \citep{honnalgere2019classification} & Center cropping, CLAHE, random flipping, and rotation & No & Modified VGG-16 
 & C-NMC \citep{Gupta2019} & $91.7\,\%$ \\  \hline
 
$2019$  \citep{prellberg2019acute} & Center cropping, random flipping, rotations, and translations & No & ResNeXt with Squeeze-and-Excitation modules & C-NMC \citep{Gupta2019}& $89.8\,\%$ \\  \hline
 
$2019$ \citep{pan2019neighborhood} & Center cropping, resizing, random rotation & No & Fine-tuned ResNet & C-NMC \citep{Gupta2019} & $92.5\,\%$ \\  \hline
\end{tabular}
\label{tab:summary_DL_methods}
\end{table*}

\subsection{Contributions}
The above discussions in Section~\ref{Literature} on the automatic ALL detection from the microscopic images recommend that different CNN-based DL methods are most widely adopted nowadays, as they alleviate the necessity of handcrafted feature extraction (see details in Table~\ref{tab:summary_ML_methods} and Table~\ref{tab:summary_DL_methods}). Although many articles have already been published, there is still room for performance improvements with better genericity of the trained model. Moreover, the CNN-based approaches experience data insufficiency to avoid overfitting, where the ensemble of different CNN architectures relieves the data scarcity limitations, as demonstrated in various articles \citep{harangi2018skin, ding2019deep, Khan2019, Liu2019, shi2019ensemble, xiao2019deepmen}. With the aforementioned thing in mind, this article intends to contribute to the exploration of building a robust ensemble model for the ALL classification, incorporating different pre-processing. We propose to aggregate the outputs of the ensemble candidate models, considering their corresponding achievements. Therefore, we proposed a weighted ensemble model, where we conduct an ablation study to determine the best weight metric. 
We perform the center-cropping of the original input images to enhance the detection results. A center-cropping enables the classifier to discover the abstract region and detailed structural information while bypassing neighboring background areas. 
Additionally, we develop five different pre-trained CNN models, such as Xception, VGG-16, DenseNet-121, MobileNet, and InceptionResNet-V2, to compare the proposed model with the same dataset and experimental settings. Our proposed weighted ensemble model outperforms the above-mentioned pre-trained models and several recently published articles on the same dataset, named  C-NMC-2019 (see details in Section~\ref{Dataset}), to our most trustworthy knowledge.   

The article's remaining sections are arranged as follows: Section \ref{materialandMethods} describes the dataset and recommended methodologies.
Section \ref{ResultsandDiscussion} reports the achieved results from various extended experiments with a proper analysis. Finally, Section \ref{Conclusion} concludes the article with future working directions.

\section{Materials and Methods}
\label{materialandMethods}
This section illustrates the materials and methodology employed for the ALL classification. Section \ref{ProposedFramework} describes the proposed framework. The utilized datasets, image pre-processing, adopted CNN architectures with ensemble techniques are explained in Sections \ref{Dataset}, \ref{Preprocessing}, and \ref{CNN}, respectively. After that, the training protocol and evaluation metrics are explained in Sections \ref{Training Protocol} and \ref{Hardware and Evaluation}, respectively. 

\subsection{Proposed Framework}
\label{ProposedFramework}
\begin{figure*}[!ht]
  \centering
  \subfloat{\includegraphics[width=16.4cm, height=4.5cm]{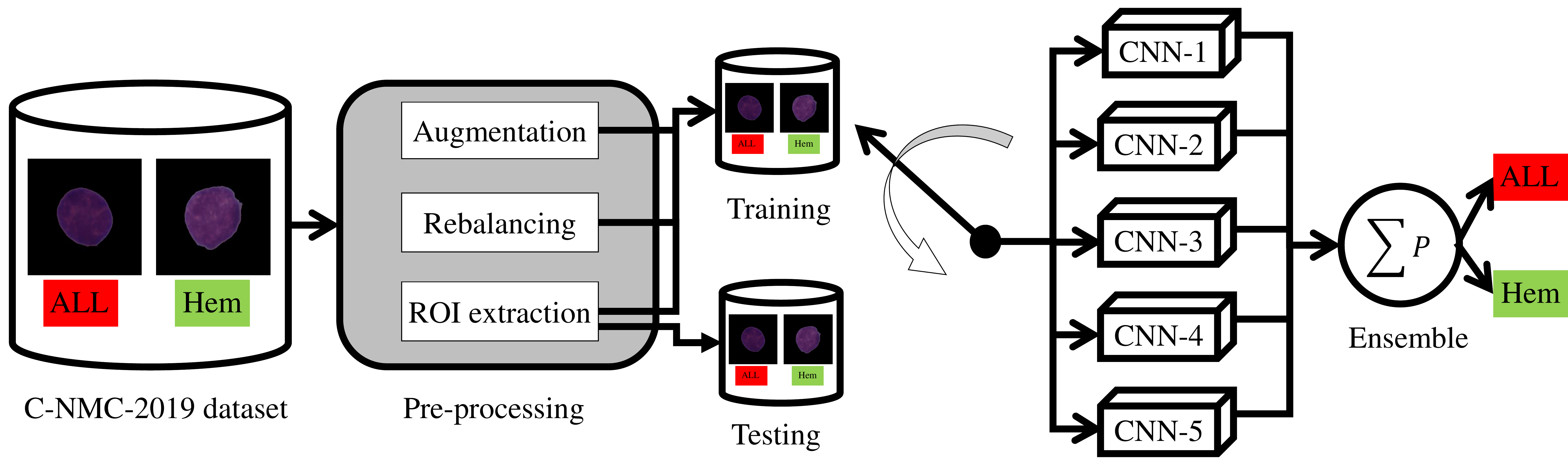}} 
  \caption{The illustration of the proposed framework, where the proposed pre-processing is the crucial integral step. The Region of Interest (ROI) is applied to both the training and testing dataset. The final categorization result is the ensemble of the different probabilities ($P$) of five different CNNs. }
  \label{block diagram}
\end{figure*}

Fig.~\ref{block diagram} represents the block exhibition of the proposed framework. The C-NMC-2019 datasets are utilized as input images for a binary ALL classification task for both training and evaluation. The image pre-processing, including rebalancing and augmentations, is integrated into the networks for the training stage. Five different well-known networks are trained with the processed images to build an ensemble classifier as it ensures better performance in the other domain of medical image classifications \citep{ding2019deep, Khan2019, shi2019ensemble}. We have adopted the pre-trained weights on the ImageNet dataset for all the networks to utilize the transfer learning policy. In the end, those five trained weights are ensembled employing soft weighted aggregation to obtain the final prediction. However, the following sections explain the essential parts of the proposed framework. 

\subsubsection{Datasets}
\label{Dataset}
\begin{figure*}[!ht]
  \centering
  \subfloat{\includegraphics[width=16.4cm, height=4cm]{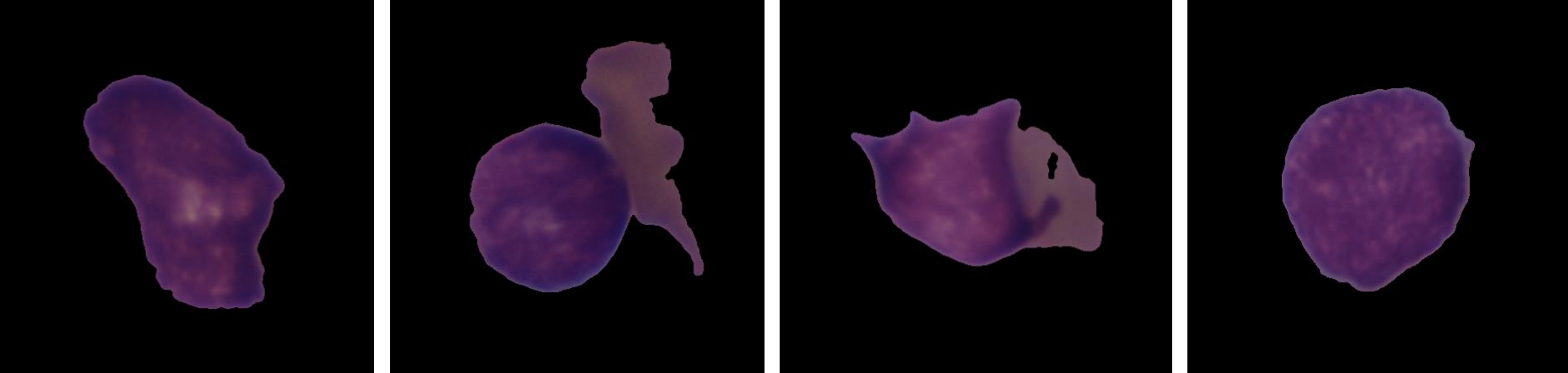}} 
  \caption{Sample images of the utilized C-NMC-2019 dataset, showing that there are unnecessary black regions around the region of interest.}
  \label{fig:dataset}
\end{figure*}

The utilized datasets were released in the medical imaging challenge, named C-NMC-2019 \citep{Gupta2019}, organized by IEEE International Symposium on Biomedical Imaging (ISBI), which contains $118$ subjects with $69$ ALL patients and $49$ Hem patients. The detailed information of the datasets is represented in Table \ref{tab:dataset}. In our proposed model, the training dataset is split into train and validation sets, and final prediction is made throughout only the preliminary test set, as shown in Table \ref{tab:dataset}. 
\begin{table*}[ht]
\centering
\footnotesize
\caption{The utilized dataset, composed of different subjects and their corresponding microscopic cell images, where we split the given training samples to obtain training and validation samples. }
\arrayrulecolor{black}
\begin{tabular}{clccccc} 
\hline
\rowcolor[rgb]{0.753,0.753,0.753} {\cellcolor[rgb]{0.753,0.753,0.753}}                         & \multicolumn{2}{c}{{\cellcolor[rgb]{0.753,0.753,0.753}}}                           & \multicolumn{2}{c}{Subjects} & \multicolumn{2}{c}{Cell Images}  \\ 
\hhline{>{\arrayrulecolor[rgb]{0.753,0.753,0.753}}--->{\arrayrulecolor{black}}----}
\rowcolor[rgb]{0.753,0.753,0.753} \multirow{-2}{*}{{\cellcolor[rgb]{0.753,0.753,0.753}}Phases} & \multicolumn{2}{c}{\multirow{-2}{*}{{\cellcolor[rgb]{0.753,0.753,0.753}}Dataset categories}} & \begin{tabular}[c]{@{}c@{}}ALL\\ (Cancerous)\end{tabular} & \begin{tabular}[c]{@{}c@{}}Hem\\ (Normal)\end{tabular}                    & \begin{tabular}[c]{@{}c@{}}ALL\\ (Cancerous)\end{tabular}    & \begin{tabular}[c]{@{}c@{}}Hem\\ (Normal)\end{tabular}                      \\ 
\hline
\multirow{2}{*}{1st}                                                                           & \multirow{2}{*}{Training samples} & Training                                                  & 32  & 19                     & 5822  & 2703                   \\ 
\cline{3-7}
                                                                                               &                        & Validation                                                & 15  & 7                      & 1450  & 686                    \\ 
\hline
2nd                                                                                            & Preliminary Test       & $-$                                                        & 13  & 15                     & 1219  & 648                     \\ 
\hline
3rd                                                                                            & Final test             & $-$                                                        & 9   & 8                      & 1761  & 825                     \\ 
\hline
\multicolumn{3}{c}{Total}                                                                                                                                                           & 69  & 49                     & 10252 & 4862                   \\
\hline
\end{tabular}
\label{tab:dataset}
\end{table*}
The resolutions of the dataset's image size are $450 \times 450$ pixels. Several sample images of the C-NMC-2019 dataset are displayed in Fig.~\ref{fig:dataset}. Table \ref{tab:dataset} shows that the dataset is imbalanced, and cancer cell images of training are around $2.15$ times more than normal cell images, making the classifier biased towards the ALL class. 
Such a biasness due to data imbalance is alleviated in the proposed framework by applying two following techniques (see details in Section~\ref{Preprocessing}). 

\subsubsection{Pre-processing}
\label{Preprocessing}
Our proposed system's crucial integral pre-processing strategies are briefly described, ensuring the better ALL prognosis system. 

Almost every image in the utilized dataset contains the region of interest in the center position with black background (see in Fig.~\ref{fig:dataset}).
Therefore, we have cropped the images centrally as the size of $300 \times 300$ pixels to decrease the overall dimension of the input data, making learning a classifier faster and easier by providing the region of interest \citep{prellberg2019acute}. 
The class imbalance is a common phenomenon in the medical imaging domain as manually annotated images are very complex and arduous to achieve \citep{harangi2018skin, hasan2021dermo}. Such a class imbalance can be partially overcome using different algorithmic-level approaches. We have used the random oversampling technique, which involves replicating the samples of minority class randomly and adding to training samples for balancing the imbalanced datasets. In our proposed model, the Hem class was oversampled to $5822$ images, and a total of $11644$ images were trained during the training process.      
Different data augmentation techniques, such as horizontal and vertical flipping, rotation, zooming, and shifting, are applied during the training process for enhancing the model's performance and building a generic model.

\subsubsection{Classifier}
\label{CNN}
As mentioned earlier in Section~\ref{Literature} that the CNN-based methods outperform ML-based methods and the radiologists with high values of balanced accuracy as proven in \citep{kermany2018identifying, rajpurkar2017chexnet}. However, single CNN may be obliquely limited when employed with highly variable and distinctive image datasets with limited samples. Transfer learning technique from a pre-trained model, which was trained on a large dataset previously, is becoming popular day by day for its advantage of using learned feature maps without having a large dataset. In this circumstance, we adopted five pre-trained networks, such as VGG-16 \citep{simonyan2014very}, Xception \citep{chollet2017xception}, MobileNet \citep{howard2017mobilenets}, InceptionResNet-V2 \citep{szegedy2017inception}, and DenseNet-121 \citep{huang2017densely} for using transfer learning application and building an ensemble classifier to categorize ALL and Hem white blood cell images.

\paragraph{\textbf{VGG-16 ($CNN_1$)}}
In $2014$, \citet{simonyan2014very} proposed a deep CNN model consisting of $16$ layers, improving the earlier AlexNet model by replacing large kernel filters.  VGG-$16$ is a deeper network (roughly twice as deep as AlexNet) by stacking uniform convolutions. The image is transferred through a stack of convolutional layers, where the filters were used with tiny receptive filters ($3\times 3$). Such a configuration allows the network to capture more excellent information with lesser computational complexity. In VGG-16, five max-pooling layers carry out spatial pooling consists of ($2 \times 2$) kernel size, which downsamples the input by a factor of $2$, bypassing the maximum value in a neighborhood of ($2 \times 2$) to the output. The VGG-$16$ ends with three fully connected layers followed by a $2$-node softmax layer.

\paragraph{\textbf{Xception ($CNN_2$)}}
Xception \citet{chollet2017xception} is an adaptation from the Inception network, replacing the Inception modules with depthwise separable convolutions. It is the introduction of CNN based network entirely with the depthwise separable convolution layers. Such a construction of the CNN model is computationally more efficient for the image classification tasks. It has $36$ convolutional layers, structured into fourteen modules, forming the feature extraction base of the network. 
The Xception top layers consist of a global average pooling layer for producing a $1 \times 2048$ vector.
The authors of the Xception network kept a number of the fully connected layer as optional as they used their model exclusively for investigating classification tasks, and that's why a logistic regression layer followed their convolution base.

\paragraph{\textbf{MobileNet ($CNN_3$)}}
In $2017$, \citet{howard2017mobilenets} proposed the MobileNet, a streamlined version of the Xception architecture, a small and low-latency CNN architecture. It also applies depthwise separable convolution for developing a lightweight deep neural network. Furthermore, MobileNet provides two parameters allowing to reduce its number of operations, which are width multiplier and resolution multiplier. The former parameter ($\alpha$) thins the number of channels, producing $\alpha \times N$ channels instead of making N channels for handling a trade-off between the desired latency and the performance. The latter channel scales the input size of the image as the MobileNet uses a global average pooling instead of a flatten. Indeed, with a global pooling, the fully connected classifier at the end of the network depends only on the number of channels, not the feature maps' spatial dimension.

\paragraph{\textbf{InceptionResNet ($CNN_4$)}}
The InceptionResNet is a deep neural network designed by \citet{he2016deep} in 2016, combining the Inception architecture \citep{szegedy2017inception} with the residual connection. 
It has a hybrid inception module inspired by the ResNet, adding the output of the convolution operation of the inception module to the input. 
In this network, the pooling operation inside the main inception modules is replaced in favor of the residual connections.

\paragraph{\textbf{DenseNet ($CNN_5$)}}
The DenseNet is a memory-saving architecture with high computational efficiency, which concatenates the feature maps of all previous layers for the inputs to the following layers \citep{ding2019deep}.
DenseNets have remarkable benefits, such as they can alleviate the vanishing gradient problem, encourage feature reuse, strengthen feature propagation,  and significantly reduce the number of parameters. 
DenseNets consists of Dense blocks, where the dimensions of the feature maps remain constant within a block, but the number of filters changes between them and Transition layers, which takes care of the downsampling, applying batch normalization, $1 \times 1$ convolution, and $2 \times 2$ pooling layers.

\paragraph{\textbf{Ensemble's Strategies}}
\citet{esteva2017dermatologist} proved that CNNs could outperform a human expert in a classification task after an exhausting learning phase on a huge annotated training set. 
However, in many cases, a sufficient number of annotated images (ground-truth) is not available, so we should improve the accuracy by other approaches. The fields of decision making and risk analysis, where information derived from several experts and aggregated by a decision-maker, have well-established literature \citep{jacobs1995methods,kittler1998combining}. 
In general, the aggregation of the opinions of the experts increases the precision of the forecast. To achieve the highest possible accuracy considering our image classification scenario, we have investigated and elaborated an automated method considering the ensemble of CNNs. 
To perform the aggregation for building an ensemble classifier, the outputs of the classification layers have been considered, which use the output of the fully-connected layers to determine probability values for each class ($n=2$). A CNN ascribes $n$ probability values $P_j \in \mathbb{R}$ to an unseen test image, where $P_j \in [0,1]$, $\forall j=1,2$, and  $\sum_{j=1}^n P_j = 1$. In ensemble modeling, we have to find out the probabilities \(P_j'\), where $P_j' \in [0,1]$, $\forall j=1,2$, and $\sum_{j=1}^n P_j' = 1$ for each test image from the probability values of the individual CNN architecture. The possible ensemble's approaches are discussed in the following paragraphs.

\paragraph{\textbf{Simple Averaging of Probabilities (SAP)}}
Averaging of the individual class confidence value is considered as one of the most commonly used ensemble model \citep{harangi2018skin, hasan2020diabetes}, which can be expressed as in Eq.~\ref{sap}. 
 \begin{equation}
 \label{sap}
   P_j' =  \frac{\sum_{k=1}^N P_{jk}}{\sum_{j=1}^n\sum_{k=1}^N P_{jk} }, \forall j=1,2.,
 \end{equation}
\noindent where \(P_{jk}\) and $N$ stand for the probability of \(CNN_k\) that a test image belongs to a particular class and the number of CNN models ($N=5$). 
Unluckily, an image may be misclassified through the SAP technique if a model with low overall accuracy treats a test image with high confidence, while the other models also provide low but non zero confidence values to the same wrong class \citep{harangi2018skin}.

\paragraph{\textbf{Weighted Ensemble of Networks (WEN)}}
To overcome the aforementioned limitations in the SAP method, we apply the weighted ensemble approach, which is an extension of  SAP where the performance of the individual network weights the contribution of each network to the final ensemble prediction.  The probabilities \(P_j'\) of each class using weighted ensemble can be derived as in Eq.~\ref{wen}.
\begin{equation}
\label{wen}
   P_j' =  \frac{\sum_{k=1}^m W_kP_{jk}}{\sum_{k=1}^m W_k }, \forall j=1,2.,
 \end{equation}
\\
\noindent where \(W_k\) denotes the weighted value of each \(CNN_k\), $\forall k \in N=5$. We have used four evaluation score, such as accuracy, AUC, F1-score, and Cohen's Kappa, as weighted values denoted as \(W_k^{acc}\), \(W_k^{auc}\), \(W_k^{f1}\), and \(W_k^{kappa}\), respectively. 
The term \(\sum_{k=1}^m W_k\) normalizes the \(P_j'\) to ensure that $P_j' \in [0,1] $, $\forall j=1,2$ and $\sum_{j=1}^nP_j' = 1$. 

\subsection{Training Policy}
\label{Training Protocol}
\begin{figure*}[!ht]
  \centering
  \subfloat{\includegraphics[width=15cm, height=5cm]{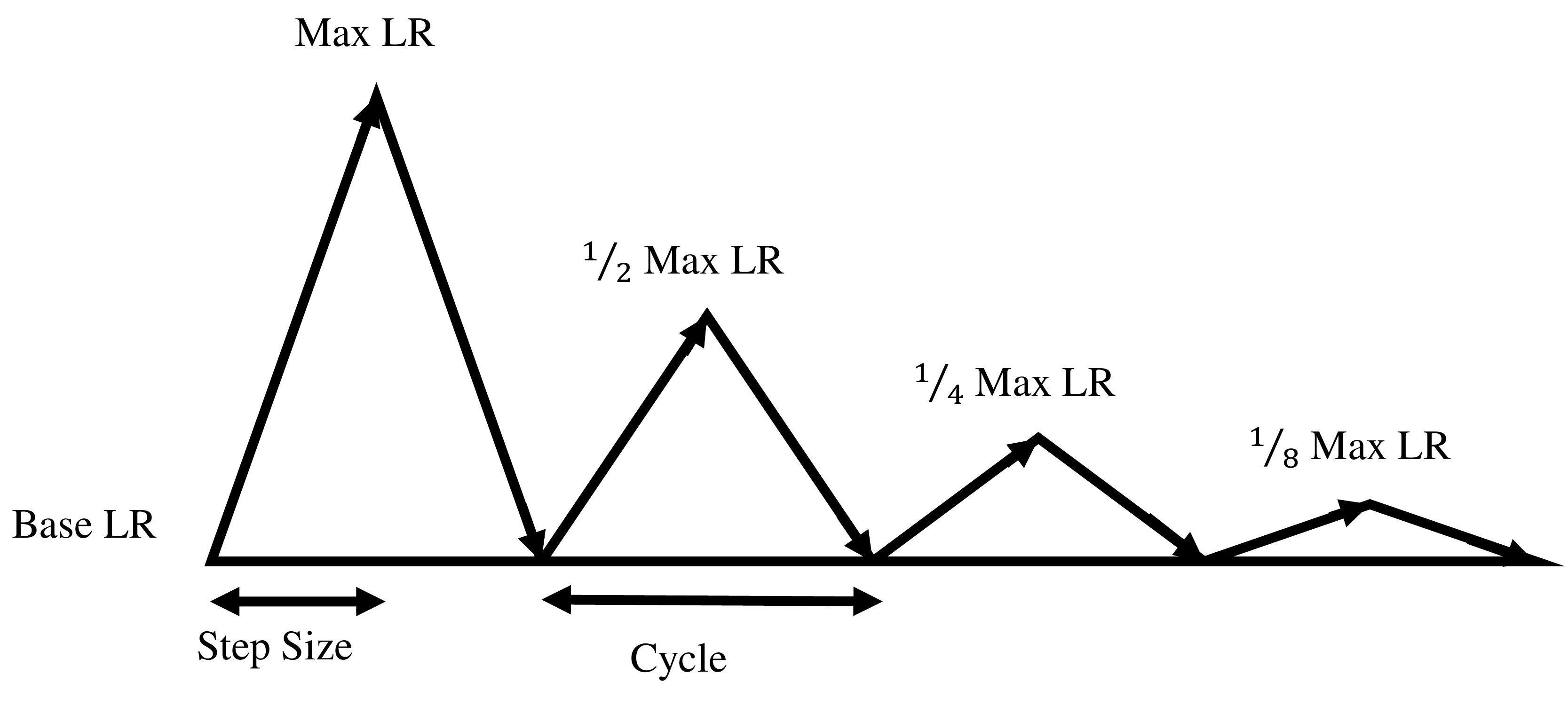}} 
  \caption{Illustration of a triangular2 type cyclic learning rate scheduler, where Base LR and Max LR are the minimum and maximum learning rate boundaries. After every cycle, the maximum learning rate is bound in half of it.}
  \label{CLR}
\end{figure*}

We employ the adamax optimizer \citep{kingma2014adam} with an initial learning rate of $0.0002$ to train all five different CNN models. The values of $\beta_1$ and $\beta_2$ are set to $0.9$ and $0.999$, respectively. Sometimes, monotonic reduction of learning rate can lead a model to stuck in either local minima or saddle points. A cyclic learning rate policy \citep{smith2017cyclical} is used for cycling the learning rate between two boundaries, such as $0.0000001$ and $0.002$. The \say{triangular2} policy shown in Fig.~\ref{CLR} is applied, and the step size is set to $StepSize = 6 \times  IterPerEpoch$, where $IterPerEpoch$ denotes the number of iterations per epoch. Categorical cross-entropy is employed as a loss function, and accuracy is chosen as the metric to train our models.  

\subsection{Hardware and Evaluation}
\label{Hardware and Evaluation}
Our proposed system is executed with the python programming language with various python and Keras APIs. The examinations are carried on a windows-10 machine with the following hardware configurations: Intel® Core™ i7-9750H CPU @ 2.60 GHz processor with Installed memory (RAM): 16GB and  NVIDIA® GeForce® GTX 1660 Ti GPU with 6 GB GDDR6 memory. 
We have utilized Weighted Precision (WP), Weighted Recall (WR), F1-score (FS), Weighted FS (WFS), Balanced Accuracy (BA), and Area Under the Curve (AUC) to evaluate the overall performance of our ALL \& Hem classifier. The following mathematical formulations describe the corresponding metric.

\begin{equation}
FS = 2 \times \frac{prcision \times recall}{prcision + recall} \nonumber  
\end{equation}

\begin{equation}
BA = \frac{specificity +  recall}{2} \nonumber 
\end{equation}

\begin{equation}
WFS = \frac{\sum_{i=0}^1n(c_i) FS(c_i)}{N}, \nonumber
\end{equation}

\noindent where \(FS(c_i)\) is the F1-score of $i^{th}$ class, \(n(c_i)\) is the number of test images in $i^{th}$ class, and N is the total number of unseen test images.

\section{Results and Discussion}
\label{ResultsandDiscussion}
This section demonstrates and interprets the obtained results from comprehensive experiments. Firstly, we explain the effect of input resolutions, such as original ($450 \times 450$) vs. center cropping ($300 \times 300$), on the training of different CNN architectures, as enlisted and described in Section~\ref{CNN}, applying various image pre-processing techniques, such as random oversampling and image augmentation. Secondly, we have aggregated the outputs of five different CNN models to enhance the ALL classification performance in terms of various evaluation metrics (see in Section~\ref{Hardware and Evaluation}). In the end, we compare our obtained results with several recent results for the same dataset and task.

\begin{figure*}[!ht]
  \centering
  \subfloat{\includegraphics[width=14cm, height=8cm]{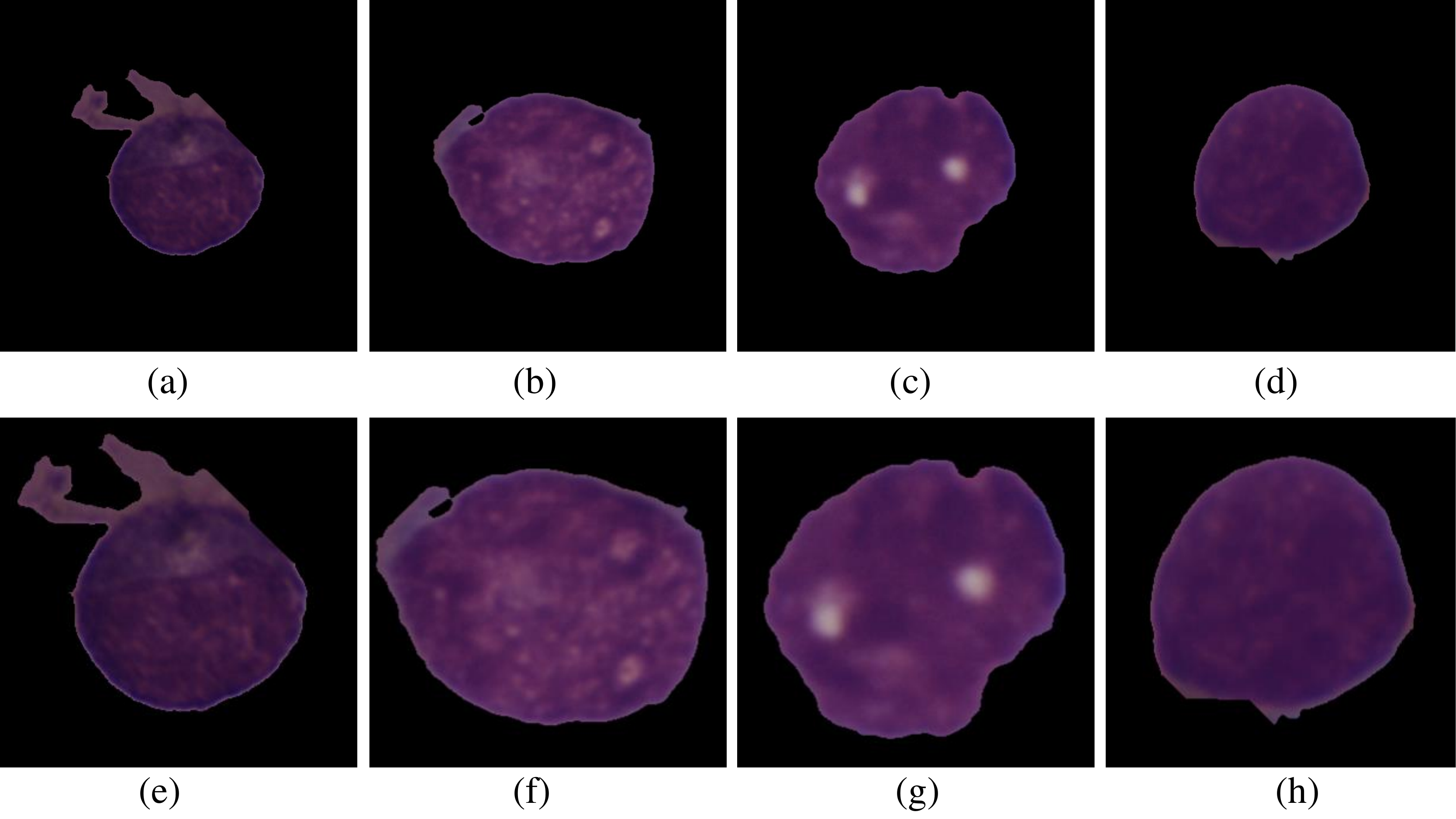}} 
  \caption{The demonstration of center-cropping of several sample images displaying no distortion of the region of interest, where (a)-(d) represents the original images with the sizes of $450 \times 450$ pixels, and (e)-(h) depicts the center-cropped images with the dimensions of $300 \times 300$ pixels. }
  \label{fig:center_crop}
\end{figure*}

\begin{figure*}[!ht]
  \centering
  \subfloat{\includegraphics[width=16.2cm, height=6cm]{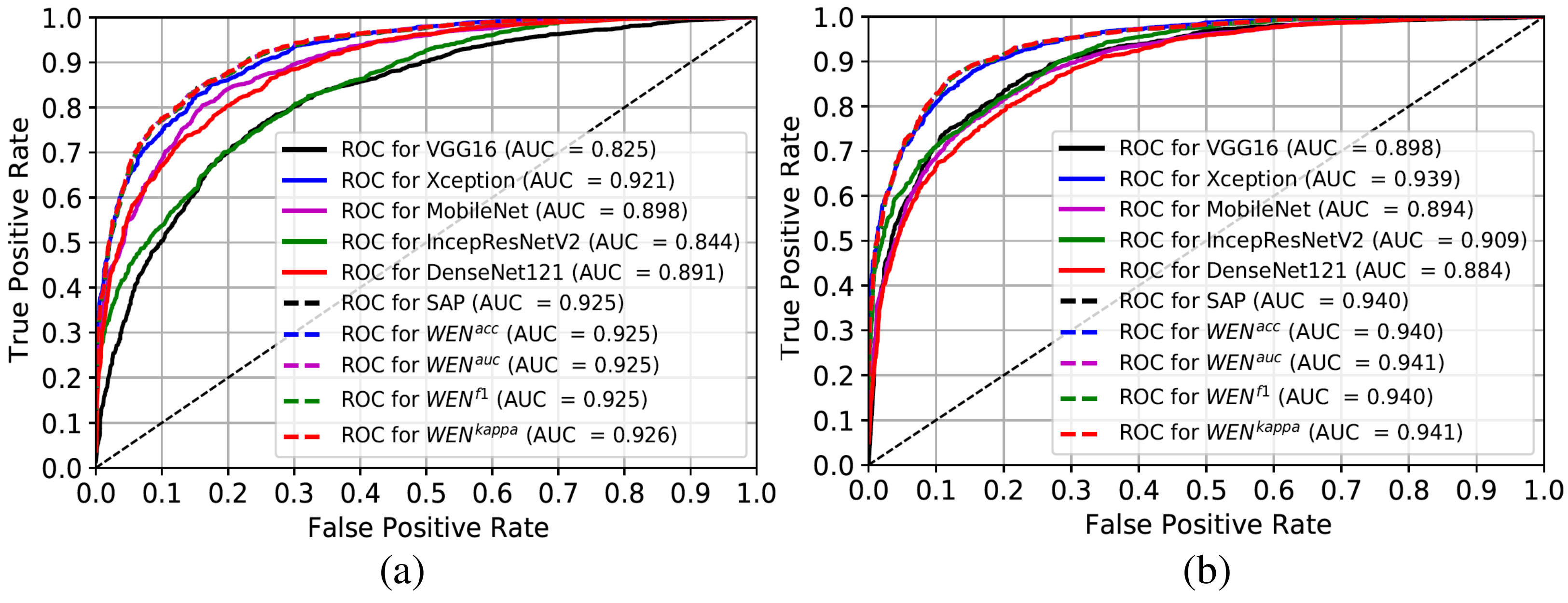}}
  \caption{The ROC curves of five different CNN models and different proposed ensemble models, where (a) for the raw input image having the size of $450 \times 450$ and (b) for the center-cropped input image with the size of $300 \times 300$.}
  \label{fig:ROC}
\end{figure*}

The sample images have been center-cropped using the nearest neighbor interpolation technique to eliminate the black border regions and provide a better area of interest, as pictorially illustrated in Fig.~\ref{fig:center_crop}. Such a center-cropping to the size of $300 \times 300$ pixels reduces the surrounded black background without distorting the original texture, shape, and other pieces of information (see in Fig.~\ref{fig:center_crop}). 
Table~\ref{tab:ClassificationResult} manifests the ALL classification results for these two different input resolutions from various classifiers, incorporating random oversampling and image augmentations. 
In both the cases of input resolutions (see in Table~\ref{tab:ClassificationResult}), it is noteworthy that the Xception model provides better results for the ALL classification. However, such a highest classification result from the Xception model is expected as it has maximum accuracy (Top-1 $79.0\,\%$) among all the available pre-trained models. 
Table~\ref{tab:ClassificationResult} demonstrates that the Xception model inputted with the sizes of $300 \times 300$ pixels has outperformed the other individual CNN models, such as VGG-16, MobileNet, InceptionResNet-V2, and DenseNet-121, outputting $86.0\,\%$-WFS, $85.9\,\%$-BA, and $93.9\,\%$-AUC. 
Those metrics beat the second-best metrics with the margins of $1.7\,\%$, $1.1\,\%$, $1.2\,\%$, $1.1\,\%$, $2.6\,\%$, and $1.8\,\%$ concerning the WP, WR, WFS, ACC, BA, and AUC, respectively.      
The experimental results (from the first ten rows in Table~\ref{tab:ClassificationResult}) also confirm that all the individual CNN model enhances their respective performance, while the center-cropped images are utilized as an input. 
The ROC curves in Fig.~\ref{fig:ROC} also reveals the benefits of the center-cropping (as $300 \times 300$), providing the higher AUCs for all the single CNN models than the original input image (as $450 \times 450$).  
Since the center-cropped images supply a better region of interest about the microscopic cell images to the networks, it empowers the CNN models to learn the most discriminating attributes, as it is experimentally validated in Table~\ref{tab:ClassificationResult}. 
\begin{table*}[!ht]
\caption{Experimental classification results on a preliminary test set representing two input image sizes of $450 \times 450$ and $300 \times 300$. The best results from each type, such as the individual CNN model and the ensemble CNN model, are depicted in bold font, whereas the second-best results are underlined for those two types.}
\footnotesize
\centering
\label{tab:ClassificationResult}
\begin{tabular}{cclcccccc} 
\hline
\rowcolor[rgb]{0.753,0.753,0.753} \multicolumn{3}{c}{Classifier}                                                                           & WP               & WR               & WFS              & ACC              & BA               & AUC               \\ 
\hline
\multirow{10}{*}{\begin{sideways}
Individual CNN models \end{sideways}} & \multirow{5}{*}
{\begin{sideways}
Original Images \end{sideways}}

 & VGG-16             & $0.775$          & $0.779$          & $0.776$          & $0.779$          & $0.744$          & $0.825$           \\ 
\cline{3-9}
                                                                                    &                                  & Xception          & $\underline{0.848}$          & $\underline{0.848}$          & $\underline{0.848}$          & $\underline{0.848}$          & $\underline{0.833}$ & $\underline{0.921}$           \\ 
\cline{3-9}
                                                                                    &                                  & MobileNet         & $0.837$          & $0.830$          & $0.820$          & $0.830$          & $0.774$          & $0.898$           \\ 
\cline{3-9}
                                                                                    &                                  & InceptionResNet-V2 & $0.784$          & $0.784$          & $0.772$          & $0.784$          & $0.722$          & $0.844$           \\ 
\cline{3-9}
                                                                                    &                                  & DenseNet-121       & $0.816$          & $0.818$          & $0.815$          & $0.818$          & $0.784$          & $0.891$           \\ 
\cline{2-9}
                                                                                    & \multirow{5}{*}{\begin{sideways}
Center-cropped \end{sideways}}  & VGG-16             & $0.843$            & $0.844$            & $0.843$         & $0.844$            & $0.825$            & $0.898$             \\ 
\cline{3-9}
                                                                                    &                                  & Xception          & $\textbf{0.865}$            & $\textbf{0.859}$            &  $\textbf{0.860}$         & $\textbf{0.859}$            & $\textbf{0.859}$            & $\textbf{0.939}$             \\ 
\cline{3-9}
                                                                                    &                                  & MobileNet         & $0.835$            & $0.837$            &  $0.835$         & $0.837$            & $0.812$            & $0.894$             \\ 
\cline{3-9}
                                                                                    &                                  & InceptionResNet-V$2$ & $\underline{0.848}$            & $\underline{0.848}$            & $0.843$         & $\underline{0.848}$            & $0.809$            & $0.909$             \\ 
\cline{3-9}
                                                                                    &                                  & DenseNet-$121$       & $0.827$            & $0.829$            & $0.826$         & $0.829$            & $0.795$            & $0.884$             \\ 
\hline
\multirow{10}{*}{\begin{sideways}
Our proposed ensemble models\end{sideways}}                                            & \multirow{5}{*}{\begin{sideways}
Original images \end{sideways}}  & SAP               & $0.859$          & $0.858$          & $0.854$          & $0.858$          & $0.820$          & $0.925$           \\ 
\cline{3-9}
                                                                                    &                                  & $WEN^{acc}$       & $0.861$          & $0.860$          & $0.855$          & $0.860$          & $0.822$          & $0.925$           \\ 
\cline{3-9}
                                                                                    &                                  & $WEN^{auc}$       & $0.861$          & $0.860$          & $0.856$          & $0.860$          & $0.823$          & $0.925$           \\ 
\cline{3-9}
                                                                                    &                                  & $WEN^{f1}$        & $0.860$          & $0.859$          & $0.854$          & $0.859$          & $0.820$          & $0.925$           \\ 
\cline{3-9}
                                                                                    &                                  & $WEN^{kappa}$     & $0.863$ & $0.862$ & $0.858$ & $0.862$ & $0.826$          & $0.926$  \\ 
\cline{2-9}
                                                                                    & \multirow{5}{*}{\begin{sideways}
Center-cropped \end{sideways}}  & SAP               & $\underline{0.886}$            & $\underline{0.886}$            & $\underline{0.884}$            & $\underline{0.886}$            & $\underline{0.859}$            & $\underline{0.940}$             \\ 
\cline{3-9}
                                                                                    &                                  & $WEN^{acc}$    & $\underline{0.886}$            & $\underline{0.886}$            & $\underline{0.884}$            & $\underline{0.886}$            & $\underline{0.859}$            & $\underline{0.940}$             \\ 
\cline{3-9}
                                                                                    &                                  & $WEN^{auc}$    & $0.885$            & $0.885$            & $0.883$            & $0.885$            & $0.856$            & $\textbf{0.941}$             \\ 
\cline{3-9}
                                                                                    &                                  & $WEN^{f1}$     & $\underline{0.886}$            & $\underline{0.886}$            & $\underline{0.884}$            & $\underline{0.886}$            & $\underline{0.859}$            & $\underline{0.940}$             \\ 
\cline{3-9}
                                                                                    &                                  & $WEN^{kappa}$  & $\textbf{0.887}$   & $\textbf{0.888}$ & $\textbf{0.886}$ & $\textbf{0.888}$ & $\textbf{0.862}$ & $\textbf{0.941}$             \\
\hline
\end{tabular}

\end{table*}

The ALL detection results have been further enhanced by proposing a weighted ensemble of those individual CNN models (see details in Section~\ref{CNN}), where the weights of the WEN are estimated from the particular model's performances, such as accuracy (acc), auc, F1-score (f1), and kappa value (kappa). The WEN models applying those weights are named as $WEN^{acc}$, $WEN^{auc}$, $WEN^{f1}$, and $WEN^{kappa}$, respectively. Four different WENs are aimed to accomplish the complete ablation studies.       
Table~\ref{tab:ClassificationResult} demonstrates the complete results of the proposed WEN and SAP to find the efficacy of WEN over the SAP techniques.  
The last ten rows of Table~\ref{tab:ClassificationResult} tell that the SAP and WEN methods exceed the former individual models comparing the same input types for single and ensemble models. 
Taking the original image of $450\times 450$ pixels, it is seen from Table~\ref{tab:ClassificationResult} that the highest Xception's results in a single model are behind the ensemble results. The SAP and WEN results outperform Xception's results with some margins for the same inputted images. Similar superiority of the SAP and WEN are noticed for the center-cropped inputs having the sizes of $300\times 300$. 
Again, it is noteworthy that the WEN methods perform better than the SAP method, comparing to all the ensemble models in Table~\ref{tab:ClassificationResult}. It is also observed that kappa value-based ensemble model ($WEN^{kappa}$) with center-cropped input has $88.7\,\%$-WP, $88.8\,\%$-WR, $88.6\,\%$-WFS, $88.8\,\%$-ACC, $86.2\,\%$-BA, and $94.1\,\%$-AUC, which outperforms all the other proposed ensemble models. Although the $WEN^{auc}$ method gains the same AUC ($94.1\,\%$) as in $WEN^{kappa}$ method for the center-cropped image, the latter method exceeds the former process in terms of other metrics (see in Table~\ref{tab:ClassificationResult}).  
For the same input resolution of $300 \times 300$ pixels, the best performing $WEN^{kappa}$ model beats the single Xception model by the margins of $2.2\,\%$, $2.9\,\%$, $2.6\,\%$, $2.9\,\%$, $0.3\,\%$, and $0.2\,\%$ concerning the WP, WR, WFS, ACC, BA, and AUC, respectively. The $WEN^{kappa}$ model, inputted with $300 \times 300$ pixels, also outperforms the $WEN^{kappa}$ model, inputted with $450\times 450$ pixels by the margins of $2.4\,\%$, $2.6\,\%$, $2.8\,\%$, $2.6\,\%$, $2.9\,\%$, and $1.5\,\%$ concerning the same metrics (serially).

Further investigation on the obtained results for two different input resolutions and various proposed and developed models are displayed in Fig.~\ref{fig:ROC}, conferring the ROC curves and their corresponding AUC values.   
Both the figures in Fig.~\ref{fig:ROC} point that all the pre-trained single models perform better when they are fine-tuned with center-cropped images with the resolution of $300 \times 300$. The pre-trained VGG-16, Xception, MobileNet, InceptionResNet-V2, and DenseNet-121 networks outperform themself by the margins of $7.3\,\%$, $1.8\,\%$, $-0.4\,\%$, $6.5\,\%$, and $-0.7\,\%$ concerning the AUC values, when trained with  center-cropped $300 \times 300$ pixels. Although the center-cropped defeats with the low margins in two cases, it wins in the other three cases with the greater margins. 
However, the proposed ensemble models' ROC curves confirm that they better the individual CNN model, whatever the input resolutions, either center-cropped or not. In both the cases of input resolutions, the proposed $WEN^{kappa}$ beats all the remaining models, providing the best ROC curve with the maximum AUC value. In the end, the proposed $WEN^{kappa}$ model has outputted the best ALL categorization results when inputted with the $300 \times 300$ pixels (center-cropped), as experimentally verified in the ROC curves in Fig.~\ref{fig:ROC}.

The detailed class-wise performances of ALL classification by the two best-performing classifiers with the center-cropped inputs, such as Xception from individual CNN and kappa-based weighted ensemble ($WEN^{kappa}$)  from the proposed fusion models, are exhibited in Table~\ref{tab:CM} (left) and Table~\ref{tab:CM} (right), respectively. 
\begin{table*}[!ht]
\centering
\caption{The confusion matrix with $1867$ unseen test samples ($1219$-ALL and $648$-Hem samples) with the resolutions of $300 \times 300$ pixels, where the left table is for individual model (Xception)  and the right table is for the proposed $WEN^{kappa}$ model. }
\label{tab:CM}
\begin{tabular}{cccclllllllcc} 
\cline{1-4}\cline{10-13}
                                                                                                       &     & \multicolumn{2}{c}{Predicted}                                                                                                                                                                                                                                & \multicolumn{1}{c}{} &  &  &  &  &                                                                                                                            &                         & \multicolumn{2}{c}{Predicted}                                                                                                                                                                                                               \\ 
\cline{3-4}\cline{12-13}
                                                                                                       &     & Hem                                                                                                                         & ALL                                                                                                                            &                      &  &  &  &  &                                                                                                                            &                         & Hem                                                                                                                 & ALL                                                                                                                   \\ 
\hhline{----~~~~~----}
\multirow{2}{*}{\begin{sideways}
Actual ~~~~ \end{sideways}} & Hem & {\cellcolor[rgb]{0.89,0.871,0.871}}\begin{tabular}[c]{@{}>{\cellcolor[rgb]{0.89,0.871,0.871}}c@{}}557\\85.96\%\end{tabular} & \begin{tabular}[c]{@{}c@{}}91\\14.04\%\end{tabular}                                                                            &                      &  &  &  &  & \multicolumn{1}{c}{\multirow{2}{*}{\begin{sideways}
Actual ~~~~ \end{sideways}}} & \multicolumn{1}{c}{Hem} & {\cellcolor[rgb]{0.82,0.8,0.8}}\begin{tabular}[c]{@{}>{\cellcolor[rgb]{0.82,0.8,0.8}}c@{}}502\\77.47\%\end{tabular} & \begin{tabular}[c]{@{}c@{}}146\\22.53\%\end{tabular}                                                                  \\ 
\hhline{~---~~~~~~---}
                                                                                                       & ALL & \begin{tabular}[c]{@{}c@{}}173\\14.19\%\end{tabular}                                                                        & {\cellcolor[rgb]{0.894,0.867,0.867}}\begin{tabular}[c]{@{}>{\cellcolor[rgb]{0.894,0.867,0.867}}c@{}}1046\\85.81\%\end{tabular} &                      &  &  &  &  & \multicolumn{1}{c}{}                                                                                                       & \multicolumn{1}{c}{ALL} & \begin{tabular}[c]{@{}c@{}}64\\5.25\%\end{tabular}                                                                  & {\cellcolor[rgb]{0.82,0.8,0.8}}\begin{tabular}[c]{@{}>{\cellcolor[rgb]{0.82,0.8,0.8}}c@{}}1155\\94.75\%\end{tabular}  \\
\hhline{----~~~~~----}
\end{tabular}
\end{table*}

Table~\ref{tab:CM} (left) depicts that out of $648$-Hem samples, $85.96\,\%$ ($557$) images are correctly recognized, whereas $14.04\,\%$ ($91$)-Hem samples are recognized as ALL type (false positive). It also discloses that,  $85.81\,\%$ ($1046$)-ALL samples are rightly classified, whereas only $14.19\,\%$ ($173$) samples are improperly classified as Hem type (false negative). 
Contrastingly, the confusion matrix of the $WEN^{kappa}$ model (see in Table~\ref{tab:CM} (right)) notes that the proposed ensemble method essentially improves the true-positive rates by a margin of $8.94\,\%$, with only $64$ ($5.25\,\%$)-ALL samples are improperly recognized as Hem (false negative).     
The discussion on the confusion matrix tells that the true-positive rate and true-negative rate are similar in the single Xception model. In contrast, those two crucial metrics in the medical diagnostic application are essentially improved by a margin of $8.94\,\%$, while we employ the proposed $WEN^{kappa}$ model.
The obtained results tell that out of a total of $1867$ samples ( 648-Hem and 1219-ALL samples), the VGG-16, Xception, MobileNet, InceptionResNet-V2, and DenseNet-121 have recognized $495\,(76.4\,\%)$, $557\,(86.0\,\%)$, $473\,(73.0\,\%)$, $442\,(68.2\,\%)$, and $442\,(68.2\,\%)$ samples as the Hem class correctly, respectively. Those values are $1080\,(88.6\,\%)$, $1046\,(85.8\,\%)$, $1089\,(89.3\,\%)$, $1141\,(93.6\,\%)$, and $1106\,(90.7\,\%)$ for the ALL class, respectively. 
Those categorization results from the proposed $WEN^{kappa}$ model are $502\,(77.5\,\%)$ and $1155\,(94.8\,\%)$ for the Hem and ALL-classes, showing the lowest false-negative rate of $5.25\,\%$ (type-II error).

\begin{figure*}[!ht]
  \centering
  \subfloat{\includegraphics[width=16.4cm, height=11.5cm]{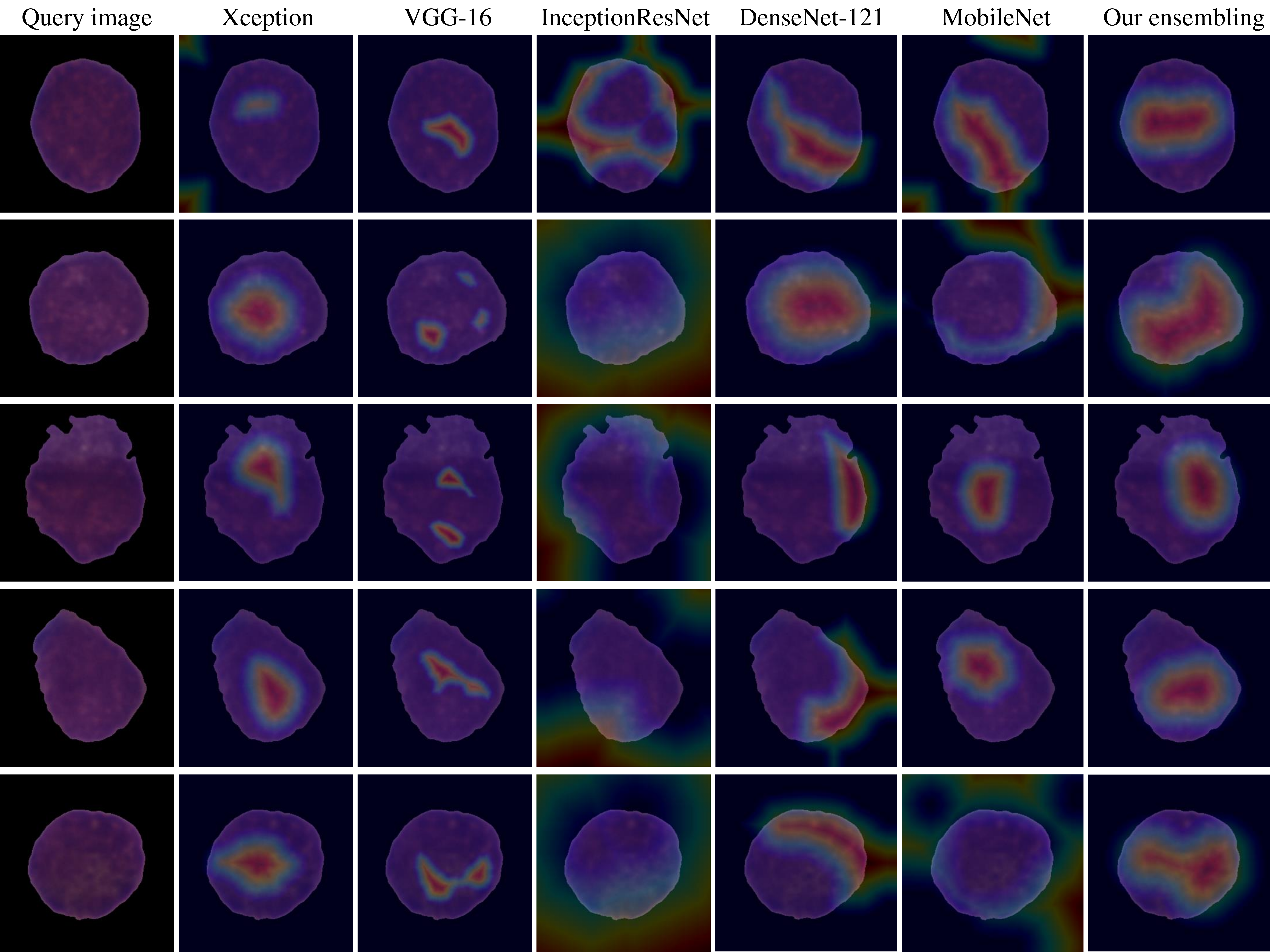}} 
  \caption{The visualization of gradient class activation maps (Grad-CAM) of the Hem class from different CNN architectures and our proposed weighted ensemble model ($WEN^{kappa}$). The Xepception, VGG-16, InceptionResNet-V2, DenseNet-121, and MobileNet fail to detect the target class, respectively, for the example in the first to fifth rows. In contrast, the proposed $WEN^{kappa}$ successfully identifies the target class in all the example cases.}
  \label{fig:Misclassified_Hem}
\end{figure*}

\begin{figure*}[!ht]
  \centering
  \subfloat{\includegraphics[width=16.4cm, height=11.5cm]{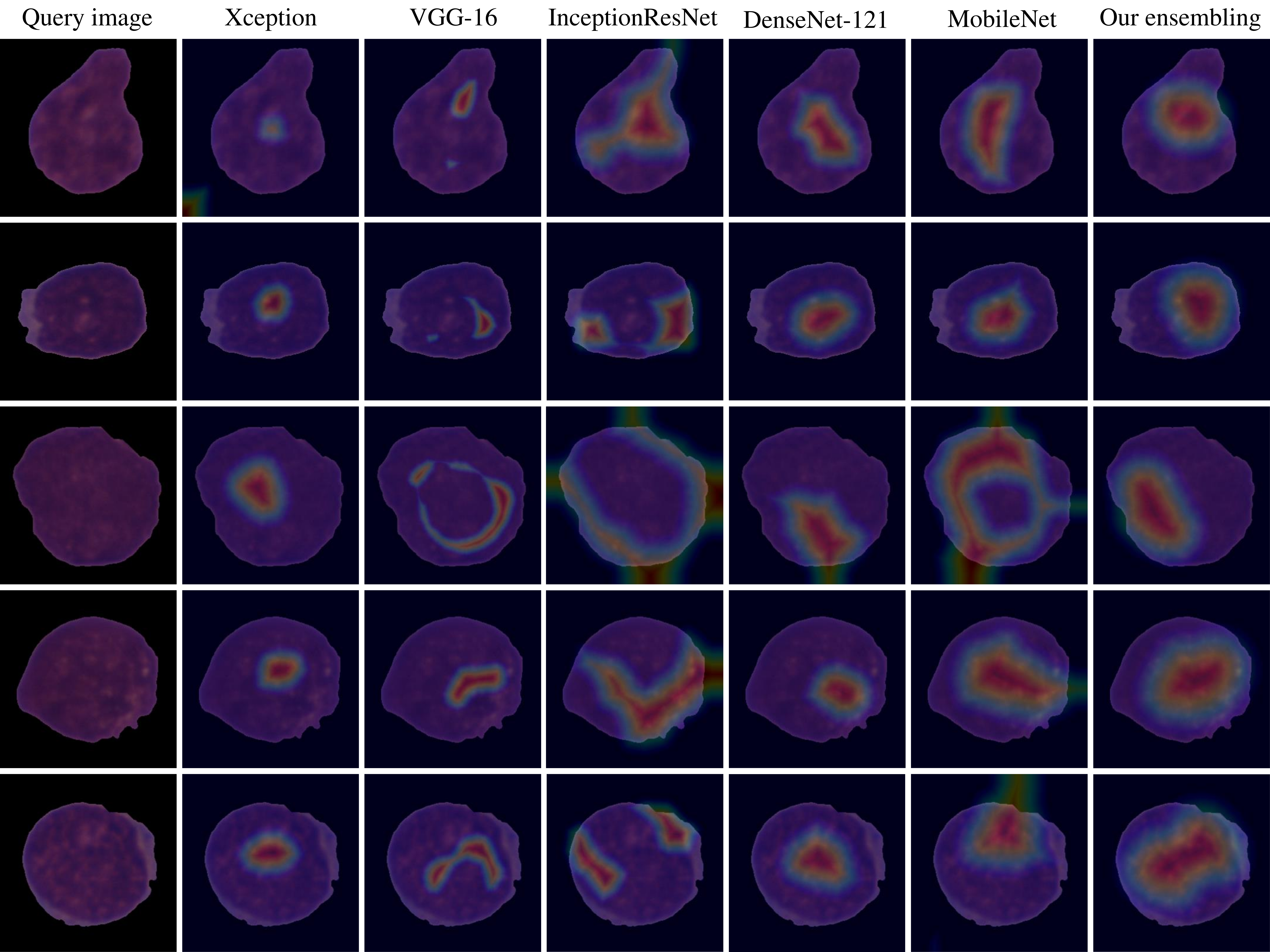}} 
  \caption{The visualization of gradient class activation maps (Grad-CAM) of the ALL class from different CNN architectures and our proposed weighted ensemble model ($WEN^{kappa}$). The Xepception, VGG-16, InceptionResNet-V2, DenseNet-121, and MobileNet fail to detect the target class, respectively, for the example in the first to fifth rows. In contrast, the proposed $WEN^{kappa}$ successfully identifies the target class in all the example cases.}
  \label{fig:Misclassified_ALL}
\end{figure*}

\begin{figure*}[!ht]
  \centering
  \subfloat{\includegraphics[width=16.4cm, height=11.5cm]{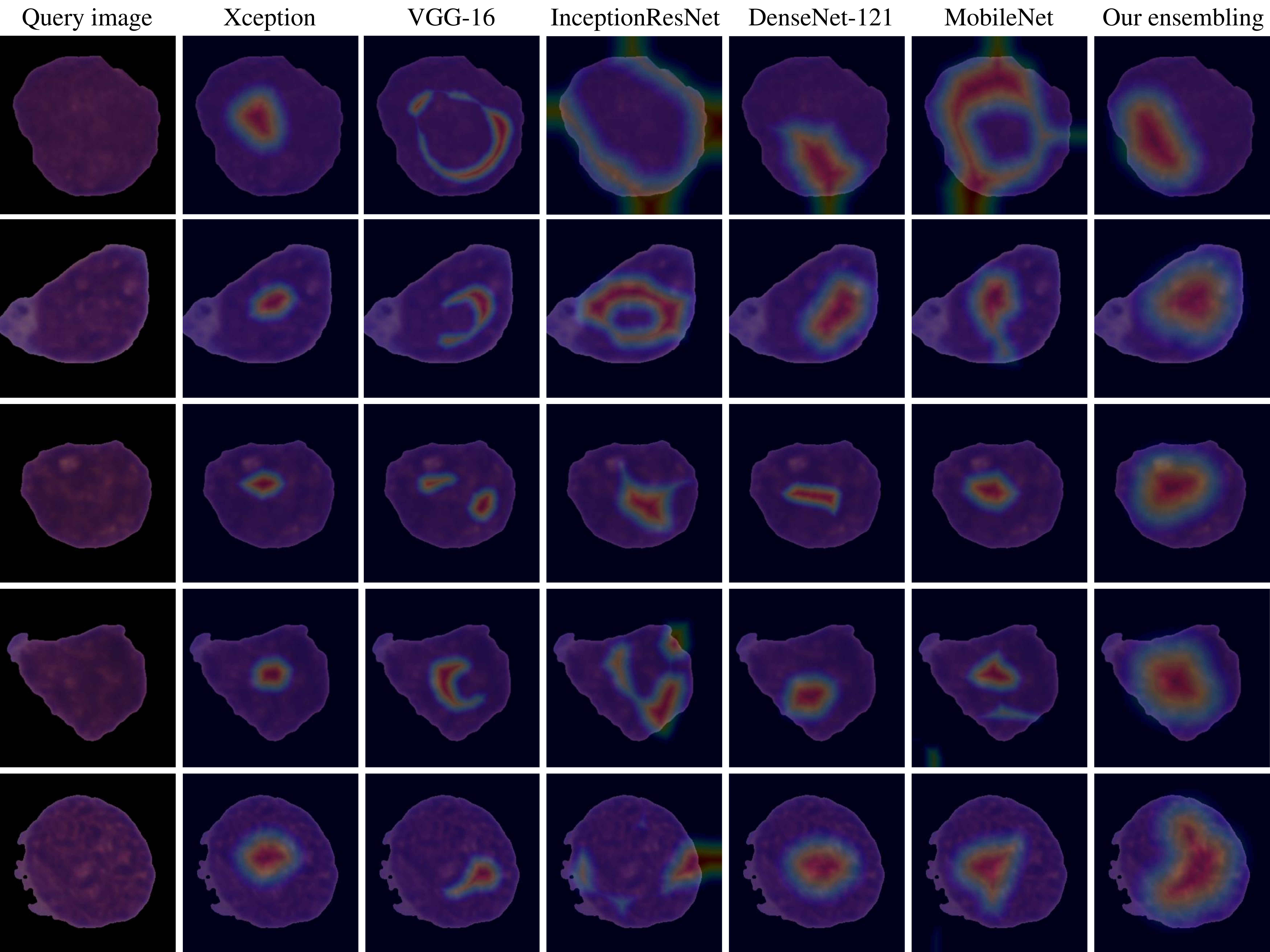}} 
  \caption{The additional visualization results for the qualitative evaluation, showing the gradient class activation maps of the ALL class from different CNN architectures and our proposed weighted ensemble $WEN^{kappa}$ model.}
  \label{fig:Grad_CAM}
\end{figure*}

For the qualitative assessment of the contribution, we present several examples of Hem and ALL samples in Fig.~\ref{fig:Misclassified_Hem} and Fig.~\ref{fig:Misclassified_ALL}, respectively,  with the class activation map overlaying (Grad-CAM), where for each instance, one of the single CNN models fails to classify. Still, our proposed best-performing $WEN^{kappa}$ model is capable of categorizing it.
The qualitative results in Fig.~\ref{fig:Misclassified_Hem} and Fig.~\ref{fig:Misclassified_ALL} expose that any single CNN may miss to recognize the target class in some examples. Still, the ensemble model successfully detects those cases, as it takes the benefits from all the candidate models to provide a final decision. It is also visible from those two figures that the Grad-CAM in the single model is coarse in most of the cases for most of the models. However, the Grad-CAM obtained from the proposed $WEN^{kappa}$ has concentrated regions in the images. For more qualitative evaluation of those concentrated Grad-CAM, additional images of the ALL class from all the single models and our $WEN^{kappa}$ are displayed in Fig.~\ref{fig:Grad_CAM}.    
However, the close inspection of all the classification results, as discussed above, concludes the superiority of the weighted ensemble techniques over the single CNN models. Such supremacy of the ensemble of methods for the same task is also proven in the earlier articles \citep{ding2019deep, Khan2019, Liu2019, shi2019ensemble} but employing a different approach with the lower outcomes.  

\begin{figure*}[!ht]
  \centering
  \subfloat{\includegraphics[width=16.4cm, height=6.8cm]{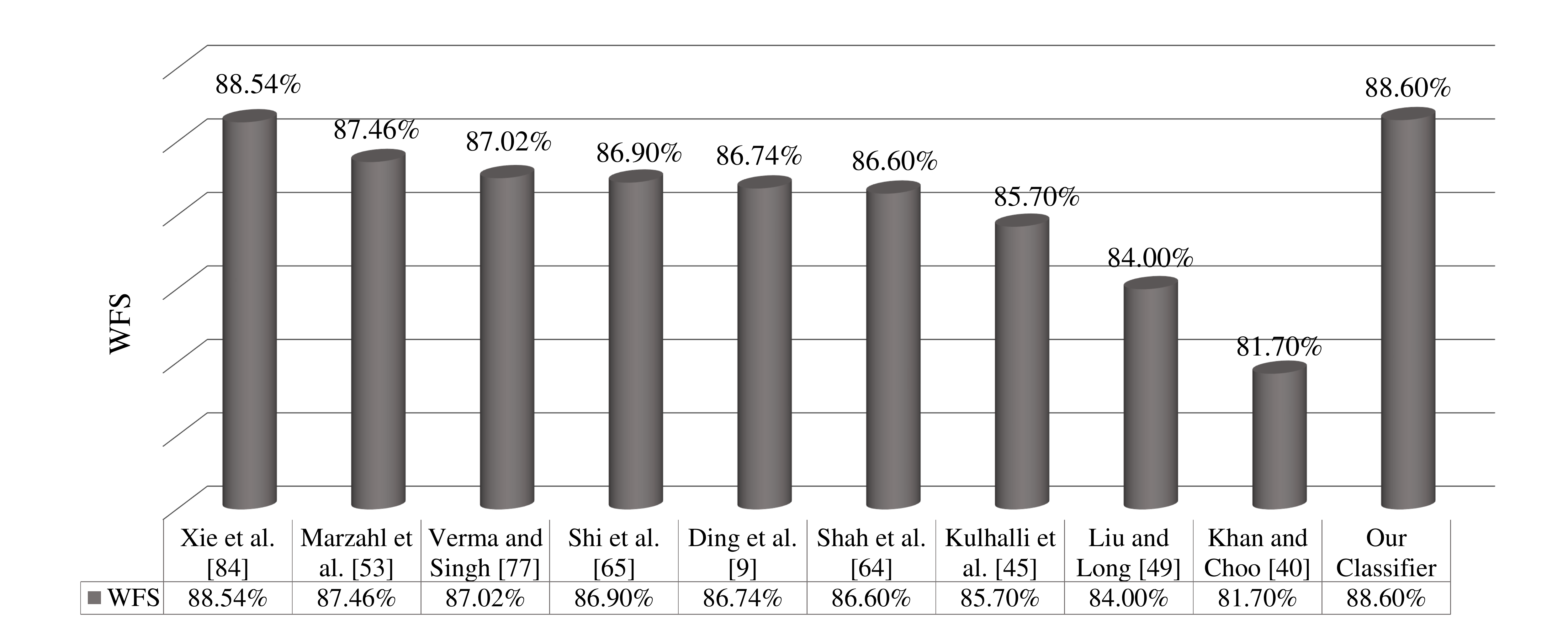}} 
  \caption{The comparison of several ALL detection methods (our proposed and recently published) on the same C-NMC dataset and the same task, showing the weighted F1-score (WFS).}
  \label{fig:performance_evaluation}
\end{figure*}

Fig.~\ref{fig:performance_evaluation} demonstrates the comparison of our proposed best performing $WEN^{kappa}$ method and other published methods on the same task and dataset, where we have shown the WFS for all the techniques. 
It is observed from Fig.~\ref{fig:performance_evaluation} that our proposed model outperforms the second-best \citep{xie2019multi} with a lower margin, but it exceeds the third-best \citep{marzahl2019classification} with a more significant margin of $1.14\,\%$. 
It is also noteworthy that our method beats the recent technique of \citet{Khan2019} with a very significant margin of $6.9\,\%$.

\section{Conclusion}
\label{Conclusion}
This article proposed and developed an automated CNN-based acute lymphoblastic leukemia detection framework for the early diagnosis, combining center-cropping, image augmentations, and class rebalancing. It was experimentally certified that the center-cropped images rather than the whole images contribute higher salient and discriminative features from the CNNs, leading to increased ALL detection. The ensemble model for the image classification with the significantly fewer training sample numbers outperforms its single candidate model. Furthermore, the weighting of the individual model in accordance with its performance enhances the aggregation results in the ensemble model. Despite the promising results of the microscopic cell image classification, it necessitates more improvement in the results, especially for the Hem class. The adversarial network can be employed to generate synthetic samples for overcoming the imbalance problems in the future. The future research direction will also focus on investigating the effect of data imbalance and accounting for the subject information fully, assuming that DL models can be adopted in more and more situations of medical image interpretation.

\section*{Author Contributions}
\textbf{C. Mondal:} Methodology, Validation, Investigation, Data Curation, Writing- Original Draft;
\textbf{M. K. Hasan:} Conceptualization, Methodology, Software, Formal analysis, Writing- Review \& Editing, Supervision;
\textbf{M. T. Jawad:}  Investigation, Writing- Review \& Editing;
\textbf{A. Dutta:}  Investigation, Writing- Original Draft;
\textbf{M. R. Islam:}  Writing- Original Draft;
\textbf{M. A. Awal:} Methodology, Writing- Review \& Editing;
\textbf{M. Ahmad:} Supervision.

\section*{Acknowledgements}
None. No funding to declare. 

\section*{Conflict of Interest}
There is no conflict of interest to publish this article.

\bibliographystyle{model2-names}

\bibliography{sample}

\end{document}